\documentclass[a4paper,12pt]{article}

\usepackage{amsmath}
\usepackage{amssymb}
\usepackage{amsthm} 
\usepackage[T1]{fontenc}
\usepackage[bf,small,tableposition=top]{caption}
\usepackage{subfig}
\usepackage{setspace}
 \usepackage{multirow}
\usepackage{graphicx}
\usepackage{epstopdf}
\usepackage{array}
\usepackage{lscape}
\usepackage{booktabs,caption,fixltx2e}
\usepackage[flushleft]{threeparttable}
\usepackage{rotating}
\usepackage{natbib}
 \usepackage{hyperref}[sectionbib, super, numbers, round]
 \hypersetup{colorlinks=true,citecolor=blue, backref=true,linktocpage=true}
\usepackage{mathtools}

\theoremstyle{plain}

\begin{document}

\begin{center}
 \Large\bf{A MCMC-type simple probabilistic approach for determining optimal progressive censoring schemes}
\end{center}
\begin{center}
\bf {\footnotesize Ritwik Bhattacharya$^1$ and Narayanaswamy Balakrishnan$^2$}
\end{center}
 \begin{center}
\textit{\scriptsize $^1$Department of Industrial Engineering, School of Engineering and Sciences, Tecnol\'{o}gico de Monterrey, Quer\'{e}taro 76130,  M\'{e}xico\\}
	
\textit{\scriptsize $^1$Department of Mathematics and Statistics, McMaster University, Hamilton, ON L8S 4K1, Canada}\\

 \end{center}

\begin{abstract}
 We present here a simple probabilistic approach for determining an optimal progressive censoring scheme by defining a probability structure on the set of feasible solutions. Given an initial solution, the new updated solution is computed within the probabilistic structure. This approach will be especially useful when the cardinality of the set of feasible solutions is large. The validation of the proposed approach is demonstrated by comparing the optimal scheme with these obtained by exhaustive numerical search.
\end{abstract}


\section{Introduction} \label{sec1}
\paragraph{}
Let us first consider the Type-II progressive censoring scheme. Suppose $n$ test units are placed on a life-testing experiment. A pre-fixed integer $m$, representing the number of failures to be observed, is chosen before the experiment starts. At the time of first failure, denoted by $X_{1:m:n}$, $R_1$ units are randomly removed from the remaining $n-1$ live units. At the time of second failure, denoted by $X_{2:m:n}$, $R_2$  units are randomly removed from the remaining  $n-R_1-2$ live units, and so on. Finally, at the time of $m$th failure, denoted by $X_{m:m:n}$, all the remaining units $n-R_1-\cdots-R_{m-1}-m$ (say, $R_m$) are removed from the testing, terminating the experiment. Therefore, the integers $R_1, \cdots, R_m$ satisfy the identity $$R_1+\cdots+R_m = n-m.$$The quantities $X_{i:m:n}, i=1, 2,\cdots,m,$ are called progressively Type-II censored ordered statistics; see  \cite{Cramer_book}. A Type-II progressive censoring scheme is characterized by the parameters $n$, $m$ and $(R_1,\cdots, R_m)$. Note that, for given $n$ and $m$, there are many choices of $R_i$'s,  and let us denote the set of all those choices by
{\footnotesize\begin{equation}\nonumber
		\mbox{CS}(n, m)=\left\{\mathcal{R}=(R_1, \cdots,R_m)\in \mathcal{N}^m~ |~ \sum_{i=1}^m R_i=n-m,~ \mathcal{N}= \{0, 1,\cdots,n-m\} \right\}.
\end{equation}}The cardinality of the set $	\mbox{CS}(n, m)$ is ${n-1 \choose m-1}$ which becomes quite large even for moderate values of $n$ and $m$. For instance, when $(n, m) = (30, 10)$, the cardinality is in fact 30045015.  Therefore, our interest is then to find the optimal Type-II progressive censoring scheme under some suitable optimality criterion function $\psi:\mbox{CS}(n, m)\rightarrow [0, \infty)$. The problem of determining optimal progressive censoring scheme has been discussed considerably in the literature; see, \citet{Cramer_book} for relevant details.  \\

\citet{Balakrishnan_book} were the first to discuss optimal progressive censoring scheme by taking $\psi$ to be the variance of the best linear unbiased estimators (BLUEs) of the model parameters and computed optimal scheme by minimizing it under normal and extreme value distributions. Due to the computational complexity, they presented the optimal schemes only up to $n=50$ and $m=3$ \citep[see][pages 197-204]{Balakrishnan_book}.  Using the same optimality criterion, \citet{Burkschat_2006, Burkschat_2007} computed optimal censoring schemes for generalized Pareto distribution. Some more choices of $\psi$, such as total time on test, expected test duration and variance of the test time were considered by \citet{Burkschat_2008}. Interestingly, in all these articles, optimal schemes mostly emerged as one-step progressive censoring, that is, censoring occurs only at one failure time (mathematically, exactly one $R_i$ is positive). \citet{Bala_2008} and \citet{Ng_2004} also introduced three choices of $\psi$ based on the asymptotic variance-covariance matrix and Fisher information, and computed optimal schemes under Weibull lifetime model. \citet{Pradhan_2009, Pradhan_2013} proposed criterion based on asymptotic variance of estimated $p$th quantile of the underlying lifetime distribution and obtained the optimal censoring schemes for generalized exponential and Birnbaum-Saunders distributions. \cite{Ritwik_2016} introduced a meta-heuristic algorithm, called variable neighborhood search (VNS), for the determination of optimal censoring schemes. They showed that the VNS algorithm performs sufficiently well for moderate to large values of $n$ and $m$. Recently, \cite{Ritwik_2020} introduced the notion of compound optimal design strategy under progressive censoring in a multi-criteria setup.  \\

In this work, we propose a method based on a probabilistic structure through which optimal censoring schemes can be computed. The proposed algorithm works in principle similar to the Markov Chain Monte Carlo (MCMC) technique; but, it is not a MCMC method as there is no target distribution. The algorithm is described in the subsequent sections along with some numerical results. For illustrative purpose, we consider a variance measure and a cost measure as optimality criterion. In fact, any optimality criteria can be utilized in the proposed method.  \\

The rest of the paper is organized as follows. The main algorithm based on the probability structure is presented in Section 2. In Section 3, the algorithmic steps under three distributions, namely, multinomial, uniform and multivariate hypergeometric distributions, are discussed. Section 4 presents some numerical results for the proposed approach and some comparative results. Finally, some concluding remarks are made in Section 5.

\section{Algorithm based on probability structure}
\paragraph{}
Suppose the members of the set $\mbox{CS}(n, m)$ can be generated through a known probability distribution on $\mathcal{R}$, say, $\pi_{\tiny{\mathcal{R}}}(\cdot)$. This is analogous to the proposal density in MCMC technique. So, in analogy to target density in MCMC, we shall construct a monotone function $f_{\tiny{\mathcal{R}|\psi(\mathcal{R})}}(\cdot)$ of criterion $\psi(\mathcal{R})$. Depending on the minimization (or maximization) of the criterion function $\psi(\mathcal{R})$, the choice of $f_{\tiny{\mathcal{R}|\psi(\mathcal{R})}}(\cdot)$ will be monotone decreasing (or increasing) function. Then, the algorithm will simulate censoring schemes $\mathcal{R}$ from $\pi_{\tiny{\mathcal{R}}}(\cdot)$ and the optimal scheme will be selected based on the following probability rule.  Let $\mathcal{R}^{\tiny{\mbox{old}}}$ be a candidate censoring scheme simulated from $\pi_{\tiny{\mathcal{R}}}(\cdot)$. A new candidate censoring scheme $\mathcal{R}^{\tiny{\mbox{new}}}$, obtained from the old candidate $\mathcal{R}^{\tiny{\mbox{old}}}$, will then be accepted with probability 
$$\alpha= \mbox{min}\left\{1, \frac{f_{\tiny{\mathcal{R}|\psi(\mathcal{R})}}(\mathcal{R}^{\tiny{\mbox{new}}}) \times \pi_{\tiny{\mathcal{R}}}(\mathcal{R}^{\tiny{\mbox{old}}})}{ f_{\tiny{\mathcal{R}|\psi(\mathcal{R})}}(\mathcal{R}^{\tiny{\mbox{old}}}) \times \pi_{\tiny{\mathcal{R}}}(\mathcal{R}^{\tiny{\mbox{new}}})    } \right\}.$$Because of the monotonic structure of $f_{\tiny{\mathcal{R}|\psi(\mathcal{R})}}(\cdot)$, the algorithm will always try to accept better candidate censoring schemes with probability $\alpha$. Consequently, starting with an initial simulated candidate $\mathcal{R}^{\tiny{\mbox{old}}}$ from 
$\pi_{\tiny{\mathcal{R}}}(\cdot)$, the algorithm will tend to reach optimal censoring scheme  after a number of simulation runs.  It is important to note that this is not a MCMC because $f_{\tiny{\mathcal{R}|\psi(\mathcal{R})}}(\cdot)$ is no longer a probability structure. In particular, we choose $$f_{\tiny{\mathcal{R}|\psi(\mathcal{R})}}(\psi(\mathcal{R})|\mathcal{R}) = e^{-\psi(\mathcal{R})}$$ for computational implementation purpose. Also, for the choice of  $\pi_{\tiny{\mathcal{R}}}(\cdot)$, we choose  multinomial, uniform and multivariate hypergeometric distributions. \\

Regarding the choices of optimality criterion $\psi(\mathcal{R})$, several options have been discussed in the literature, and a list of these options can be found in \cite{Ritwik_2020}. In this work, we consider two optimality criteria. The first one is based on a variance measure given by
$$\psi(\mathcal{R}) = \int_{0}^{1} \mbox{Var}[\ln \hat{X}_s]\mbox{d}s,$$where $\hat{X}_s$ represents the maximum likelihood estimate of $X_s$, the $s$th quantile of the underlying lifetime distribution. The aggregate variance of the
log-quantile estimates over all quantile points get summarized under this measure. Moreover, the criterion is scale-invariant meaning that the optimal scheme computed by optimizing the criterion does not change if the time scale gets changed. This is a desirable property in an optimizing criterion used in life-testing experiments \citep[see][]{Ritwik_2016}. The second criterion is based on a cost measure given by
$$\psi(\mathcal{R}) = C_{\scriptsize{\mbox{o}}} + C_{\scriptsize{\mbox{f}}}m+C_{\scriptsize{\mbox{t}}}E[X_{m:m:n}],$$where $C_{\scriptsize{\mbox{f}}}$ is the cost per unit of failure, $C_{\scriptsize{\mbox{t}}}$ is the cost per unit of duration of testing, and $C_{\scriptsize{\mbox{o}}}$ is an overall fixed cost that does not depend on the design parameters of the test. This criterion is also scale-invariant.\\

A Weibull lifetime model with distribution function of the form
\begin{equation}\nonumber
	F(x; \beta, k) = 1 - e^{-(kx)^{\beta}},
\end{equation}where $\beta>0$ and $k>0$ are the shape and scale parameters, respectively, is considered for illustrative purpose. The required expressions of $\psi(\mathcal{R})$ under Weibull model are presented in the Appendix.

\section{Different choices for $\pi_{\tiny{\mathcal{R}}}(\cdot)$}
\paragraph{}
As mentioned in the last section, $\pi_{\tiny{\mathcal{R}}}(\cdot)$ needs to be specified with some known probability distribution, and then the censoring schemes will be simulated from this distribution. A complicated probability structure may increase the computational complexity of the algorithm. Here, we choose three known probability distributions for $\pi_{\tiny{\mathcal{R}}}(\cdot)$ and then assess the corresponding numerical results.

\subsection{Multinomial distribution}
\paragraph{}
Let us assume a multinomial distribution with probability mass function 
\begin{equation}\label{MN}
	\pi_{\tiny{\mathcal{R}}}(R_1, \cdots, R_m; p_1, \cdots, p_m) = \frac{(n-m)!}{R_1!\cdots R_m!}p_1^{R_1} \cdots p_m^{R_m},
\end{equation}where $(n-m)$ and $p_i>0, i=1,\cdots, m,$ are the parameters; see \citet{Johnson_book} for various properties of this distribution. The simulation of a candidate from (\ref{MN}) is straightforward and any candidate generated from (\ref{MN}) automatically would satisfy the censoring constraint imposed by progressive censoring, namely, $R_1+\cdots+R_m=n-m$. Now, to run the algorithm, we adopt the following two steps:
\begin{description}
	\item[Step 1 (Simulating $\mathcal{R}_{0}$)] An initial candidate $\mathcal{R}_{0}$ can be simulated in the following manner. Generate $m$ random numbers $u_i, i=1,\cdots, m,$ from a uniform distribution $U[0, 1]$, and set $$p_i = \frac{u_i}{\sum_{i=1}^{m} u_i} .$$Then, $\mathcal{R}_0$ can be directly simulated from the multinomial distribution with parameters $n-m$ and $(p_1, p_2,\cdots, p_m)$;
	\item[Step 2 (Simulating $\mathcal{R}_1$ from $\mathcal{R}_0$)] Now, given $\mathcal{R}_0$, a new $\mathcal{R}_1$ can be randomly generated from $\mathcal{R}_0$ in the following manner. We randomly replace full/partial components of $\mathcal{R}_0$ with observations generated from a multinomial distribution. First, randomly choose $m_1 (0<m_1\leq m)$ positions of $\mathcal{R}_0$. Let $(R_1^{'}, \cdots,$ $ R_{m_1}^{'})$ and $(p_{R_1^{'}},\cdots, R_{m_1}^{'})$ denote the corresponding positions and probability values in $\mathcal{R}_0$, respectively. Now, we recalculate the probabilities for these $m_1$ positions as $$q_j = \frac{p_{R_j^{'}}}{\sum_{j=1}^{m_1} p_{R_j^{'}}},$$where $j=1,\cdots, m_1$. We then simulate $m_1$ values from the multinomial distribution with parameters $\sum_{j=1}^{m_1} R_j^{'}$ and $(q_1, \cdots, q_{m_1})$, and concatenate with $\mathcal{R}_0$.
\end{description}
Following the above two steps, a new candidate will be accepted with probability $$\mbox{min}\left\{1, \frac{f_{\tiny{\mathcal{R}|\psi(\mathcal{R})}}(\psi(\mathcal{R}_1)|\mathcal{R}_1) \times \pi_{\tiny{\mathcal{R}}}(\mathcal{R}_0; p_1, p_2,\cdots, p_m)}{  f_{\tiny{\mathcal{R}|\psi(\mathcal{R})}}(\psi(\mathcal{R}_0)|\mathcal{R}_0) \times \pi_{\tiny{\mathcal{R}}}(\mathcal{R}_1; p_1, p_2,\cdots, p_m)   } \right\}.$$At each time of acceptance of a candidate, a new updated optimal solution is obtained, and we would therefore expect to obtain the optimal solution after performing a number of simulation runs.

\subsection{Uniform distribution}
\paragraph{}
Let us consider a discrete uniform distribution to construct $\mathcal{R}_0$, $\mathcal{R}_1$ and $\pi_{\tiny{\mathcal{R}}}(\cdot)$ in the following manner. First, generate a random number between 0 and $n-m$ and set it as $R_1$. Then, generate another number between 0 and $n-m-R_1$ and set it as $R_2$. Continue the process until all $m$ positions are filled up. Thus, $\mathcal{R}_0$ gets generated and
\begin{equation*}
	\pi_{\tiny{\mathcal{R}}}(\mathcal{R}_0) = \frac{1}{(n-m)(n-m-R_1)\cdots(n-m-R_1-\cdots -R_{m-1})}.
\end{equation*}Next, we generate a new $\mathcal{R}_1$ from $\mathcal{R}_0$ as follows. Randomly choose a position in $\mathcal{R}_0$ and let the corresponding value at that position be $r_0$. Then, first generate a number randomly between 0 and $n-m-r_0$ and set it as $R_1$ in $\mathcal{R}_1$; generate another number between 0 and $n-m-R_1$ and set it as $R_2$ in $\mathcal{R}_1$; and continue this process until all $m$ positions are filled up. Thus, we can generate $\mathcal{R}_1$ and
\begin{equation*}
	\pi_{\tiny{\mathcal{R}}}(\mathcal{R}_1) = \frac{1}{(n-m-r_0)(n-m-R_1)\cdots(n-m-R_1-\cdots-R_{m-1})}.
\end{equation*}Thence, a new candidate will be accepted with probability
$$\mbox{min}\left\{1, \frac{f_{\tiny{\mathcal{R}|\psi(\mathcal{R})}}(\psi(\mathcal{R}_1)|\mathcal{R}_1) \times \pi_{\tiny{\mathcal{R}}}(\mathcal{R}_0)}{  f_{\tiny{\mathcal{R}|\psi(\mathcal{R})}}(\psi(\mathcal{R}_0)|\mathcal{R}_0) \times \pi_{\tiny{\mathcal{R}}}(\mathcal{R}_1)   } \right\}.$$

\subsection{Multivariate hypergeometric distribution}
\paragraph{}
Let us assume a multivariate hypergeometric distribution with probability mass function
\begin{equation}\label{MHG}
	\pi_{\tiny{\mathcal{R}}}(R_1,\cdots, R_m; M, (M_1,\cdots, M_m), R) = \frac{\binom{M_1}{R_1}\cdots\binom{M_m}{R_m}}{\binom{M}{R}},
\end{equation}where $M$, $(M_1,\cdots, M_m)$ and $R$ are the parameters with $M_j = n-m,$ for $j =1,\cdots, m,$ $M=\sum_{j=1}^m M_j = m(n-m)$ and $R=\sum_{j=1}^m R_j = n-m$; see \citet{Johnson_book} for properties of this distribution. One may readily note that any candidate simulated from (\ref{MHG}) would automatically satisfy the constraint imposed by progressive censoring. For given $n$ and $m$, simulation of one candidate as $\mathcal{R}_0$ is straightforward. Next, we use the following steps to generate $\mathcal{R}_1$ from $\mathcal{R}_0$. First, randomly choose $m_1 (0<m_1\leq m)$ positions of $\mathcal{R}_0$, and let $(R_1^{'}, \cdots,$ $ R_{m_1}^{'})$ denote the corresponding positions. Next, simulate a candidate of length $m_1$ from a multivariate hypergeometric distribution with parameters $M_j=\sum_{j=1}^{m_1}R_j^{'}$, $M=\sum_{j=1}^{m_1} M_j$ and $R=\sum_{j=1}^{m_1}R_j^{'}$, and concatenate with $\mathcal{R}_0$. Then, proceeding as before, a new candidate will be accepted with probability 
$$\mbox{min}\left\{1, \frac{f_{\tiny{\mathcal{R}|\psi(\mathcal{R})}}(\psi(\mathcal{R}_1)|\mathcal{R}_1) \times \pi_{\tiny{\mathcal{R}}}(\mathcal{R}_0;M, (M_1,\cdots, M_m), R)}{  f_{\tiny{\mathcal{R}|\psi(\mathcal{R})}}(\psi(\mathcal{R}_0)|\mathcal{R}_0) \times \pi_{\tiny{\mathcal{R}}}(\mathcal{R}_1; M, (M_1,\cdots, M_m), R)   } \right\}.$$

\section{Numerical results}
\paragraph{}
We shall use Weibull distribution with parameters $(\beta, k)$=(1, 0.5), (1, 1) and (1, 2) as the underlying lifetime model. We evaluate the performance of the proposed algorithm by comparing it with the optimal solution obtained through exhaustive search. Due to the large cardinality of the set $\mbox{CS}(n, m)$, exhaustive search is only possible for small value of $n$ and $m$. For this reason, we set $(n, m)$=(10, 5), (15, 5), (20, 5) and compared the optimal solutions by a relative efficiency measure defined as
{\footnotesize{\begin{equation*}
	\mbox{R.eff}_1 = \frac{\mbox{The value $\psi(\mathcal{R}^{*})$ corresponding to optimal scheme $\mathcal{R}^{*}$ in exhaustive search}}{\mbox{The value $\psi(\mathcal{R}^{\bullet})$ corresponding to optimal scheme $\mathcal{R}^{\bullet}$ in proposed approach}}.
\end{equation*}  }}The optimal schemes and the corresponding relative efficiencies so determined are presented in Table 1. These results show that even though the proposed approach does not achieve the true optimal scheme $\mathcal{R}^{*}$, but the corresponding relative efficiencies are very close to 1. For this reason, we refer to the obtained solution $\mathcal{R}^{\bullet}$ as near-optimal censoring scheme. In addition, we have reported two quantities $n_{\tiny{\mbox{it}}}$ and $n_{\tiny{\mbox{ac}}}$ representing the number of simulation runs and accepted candidates, respectively. The corresponding results show the performance of multinomial distribution to be better than the uniform and multivariate hypergeometric distributions in that multinomial needs less number of accepted candidates than the other two distributions to reach near-optimal scheme with the same relative efficiency when the cardinality of the set $\mbox{CS}(n, m)$ is large.  \\

As pointed out in \cite{Ritwik_2016}, the VNS algorithm provides near-optimal censoring schemes for large values of $n$ and $m$. Therefore, we also compare the performance of the proposed approach with the VNS algorithm, and the corresponding results are reported in Table 2. We use the notation $\mathcal{R}^{\dagger}$ to denote the near-optimal scheme obtained through the VNS algorithm, and we define a relative efficiency measure as
\begin{equation*}
			\mbox{R.eff}_2 = \frac{\begin{aligned}
				\mbox{The value of $\int_{0}^{1} \mbox{Var}[\ln \hat{X}_s]\mbox{d}s$ corresponding to optimal}\\
				\mbox{scheme $\mathcal{R}^{\dagger}$ in VNS algorithm}\end{aligned}
			}{\begin{aligned}\mbox{
				The value of $\int_{0}^{1} \mbox{Var}[\ln \hat{X}_s]\mbox{d}s$ corresponding to optimal}\\
			\mbox{ scheme $\mathcal{R}^{\bullet}$ in proposed approach}\end{aligned}}.
\end{equation*}
%{\footnotesize{\begin{equation*}
%			\mbox{R.eff}_2 = \frac{\mbox{The value of $\int_{0}^{1} \mbox{Var}[\ln \hat{X}_s]\mbox{d}s$ corresponding to optimal scheme $\mathcal{R}^{\dagger}$ in VNS algorithm}}{\mbox{The value of $\int_{0}^{1} \mbox{Var}[\ln \hat{X}_s]\mbox{d}s$ corresponding to optimal scheme $\mathcal{R}^{\bullet}$ in proposed approach}}.
%\end{equation*}  }}
The results in Table 2 show that, even though the near-optimal schemes are different, the corresponding $\mbox{R.eff}_2$ are always 1 indicating that   the proposed approach performs well as compared to the VNS algorithm. While reporting the optimal schemes, we use the notation $a^b$ to denote that $a$ is repeated $b$ times. For example, the scheme $((0^5, 20, 0^4))$ for the case $(n, m)=(30, 10)$ under VNS algorithm in Table 2 refers to the scheme $(0, 0, 0, 0, 0, 20, 0, 0, 0, 0)$.  In a similar vein, the relative efficiency under cost criterion measure is defined as  
%{\footnotesize{\begin{equation*}
%			\mbox{R.eff}_3 = \frac{\mbox{The value of $C_{\scriptsize{\mbox{o}}} + C_{\scriptsize{\mbox{f}}}m+C_{\scriptsize{\mbox{t}}}E[X_{m:m:n}]$ corresponding to optimal scheme $\mathcal{R}^{\dagger}$ in VNS algorithm}}{\mbox{The value of $C_{\scriptsize{\mbox{o}}} + C_{\scriptsize{\mbox{f}}}m+C_{\scriptsize{\mbox{t}}}E[X_{m:m:n}]$ corresponding to optimal scheme $\mathcal{R}^{\bullet}$ in proposed approach}}.
%\end{equation*}  }}
\begin{equation*}
			\mbox{R.eff}_3 = \frac{\begin{aligned}\mbox{The value of $C_{\scriptsize{\mbox{o}}} + C_{\scriptsize{\mbox{f}}}m+C_{\scriptsize{\mbox{t}}}E[X_{m:m:n}]$ corresponding to}\\
			\mbox{ optimal scheme $\mathcal{R}^{\dagger}$ in VNS algorithm}\end{aligned}
					}{\begin{aligned}\mbox{The value of $C_{\scriptsize{\mbox{o}}} + C_{\scriptsize{\mbox{f}}}m+C_{\scriptsize{\mbox{t}}}E[X_{m:m:n}]$ corresponding to}\\
						\mbox{ optimal scheme $\mathcal{R}^{\bullet}$ in proposed approach}\end{aligned}}.
\end{equation*}Although, the numerical results are not reported here, it is found that the efficiencies $\mbox{R.eff}_3$ are all once again close to one.

\begin{table}[h]
	\centering
	\tiny
	\caption{Comparison between the optimal censoring schemes obtained through exhaustive search and the proposed algorithm under the criterion $\psi(\mathcal{R}) = \int_{0}^{1} \mbox{Var}[\ln \hat{X}_s]\mbox{d}s$ in the case of Weibull lifetime model.}
	\begin{tabular}{llllllllll}\toprule
%				\multicolumn{9}{c}{Weibull parameters, $(\beta, k) = (0.5, 1)$}\\\toprule
Parameters&		\multirow{2}{*}{$(n, m)$} && \multicolumn{2}{c}{Exhaustive search} && \multicolumn{4}{c}{Multinomial}\\\cline{4-5}\cline{7-10}
$(\beta, k)$&		&& $\mathcal{R}^{*}$ & $\psi(\mathcal{R}^{*})$ && $n_{\tiny{\mbox{it}}}$  & $n_{\tiny{\mbox{ac}}}$ & $\mathcal{R}^{\bullet}$ & R.eff$_1$ \\\midrule
\multirow{5}{*}{(0.5, 1)}&		(10, 5) && 	(0, 4, 1, 0, 0) & 2.4261 &&  500 & 500 & (0, 4, 0, 0, 1) &   0.9995      \\        
&		&&                             &                 && 1000 & 1000 & (0, 4, 0, 0, 1) & 0.9995\\
&		&&                             &                 && 10000 & 8902 & (0, 4, 0, 0, 1) & 0.9995\\
&		&&                             &                 && 50000 & 18936 & (0, 4, 0, 0, 1) & 0.9995\\
&		&&                             &                 && 100000 & 99952 & (0, 4, 0, 0, 1) & 0.9995\\\bottomrule
Parameters &		\multirow{2}{*}{$(n, m)$} && \multicolumn{2}{c}{Exhaustive search} && \multicolumn{4}{c}{Uniform}\\\cline{4-5}\cline{7-10}
$(\beta, k)$ &		&& $\mathcal{R}^{*}$ & $\psi(\mathcal{R}^{*})$ && $n_{\tiny{\mbox{it}}}$  & $n_{\tiny{\mbox{ac}}}$ & $\mathcal{R}^{\bullet}$  & R.eff$_1$ \\\midrule	             
\multirow{5}{*}{(0.5, 1)}&		(10, 5) && 	(0, 4, 1, 0, 0) & 2.4261 && 500 & 412 & (0, 4, 0, 0, 1) & 0.9990\\	             
&		&&                             &                 && 1000 & 420 & (0, 4, 0, 0, 1) &  0.9990\\
&		&&                             &                 && 10000 & 9508 & (0, 4, 0, 0, 1) &  0.9990\\
&		&&                             &                 && 50000 &  47487 & (0, 4, 0, 0, 1) &  0.9990\\
&		&&                             &                 &&  100000 &   49685         &   (0, 4, 0, 0, 1) & 0.9990\\\bottomrule
Parameters &		\multirow{2}{*}{$(n, m)$} && \multicolumn{2}{c}{Exhaustive search} && \multicolumn{4}{c}{Multivariate hypergeometric}\\\cline{4-5}\cline{7-10}
$(\beta, k)$ &		&& $\mathcal{R}^{*}$ & $\psi(\mathcal{R}^{*})$ && $n_{\tiny{\mbox{it}}}$  & $n_{\tiny{\mbox{ac}}}$ & $\mathcal{R}^{\bullet}$  & R.eff$_1$ \\\midrule	  
\multirow{5}{*}{(0.5, 1)} &		(10, 5) && 	(0, 4, 1, 0, 0) & 2.4261 && 500 & 358 & (1, 4, 0, 0, 0) & 0.9995\\
&		&&                                &                 && 1000 & 600 & (1, 4, 0, 0, 0) & 0.9995\\         
&		&& 	                            &                 && 10000 & 7016 & (1, 1, 0, 2, 1) & 	0.9995\\              
&		&&                                &                 && 50000 &  35566 & (1, 1, 1, 0, 2) & 0.9998\\
&		&&                                &                 && 100000 &    26463      &       (0, 0, 6, 3, 6)                 & 0.9998\\\bottomrule
Parameters &		\multirow{2}{*}{$(n, m)$} && \multicolumn{2}{c}{Exhaustive search} && \multicolumn{4}{c}{Multinomial}\\\cline{4-5}\cline{7-10}
$(\beta, k)$&		                 && $\mathcal{R}^{*}$ & $\psi(\mathcal{R}^{*})$ && $n_{\tiny{\mbox{it}}}$   & $n_{\tiny{\mbox{ac}}}$ & $\mathcal{R}^{\bullet}$ & R.eff$_1$ \\\midrule
\multirow{5}{*}{(1, 1)} &	(15, 5) && 	(0, 10, 0, 0, 0) & 0.4983 &&  500 & 310 & (0, 0, 5, 5, 0) &   0.9996      \\        
&	             &&                             &                 && 1000 & 508 & (0, 7, 2, 0, 1) & 0.9996\\
&	             &&                             &                 && 10000 & 5438 & (0, 8, 1, 1, 0) & 0.9996\\
&	             &&                             &                 && 50000 & 26141 & (0, 9, 1, 0, 0) & 0.9996\\
&	             &&                             &                 && 100000 & 45858 & (0, 12, 3, 0, 0) & 0.9998\\\bottomrule
Parameters &	\multirow{2}{*}{$(n, m)$} && \multicolumn{2}{c}{Exhaustive search} && \multicolumn{4}{c}{Uniform}\\\cline{4-5}\cline{7-10}
$(\beta, k)$ & && $\mathcal{R}^{*}$ & $\psi(\mathcal{R}^{*})$ && $n_{\tiny{\mbox{it}}}$  & $n_{\tiny{\mbox{ac}}}$ & $\mathcal{R}^{\bullet}$  & R.eff$_1$ \\\midrule	             
\multirow{5}{*}{(1, 1)} &	(15, 5) && 	(0, 10, 0, 0, 0) & 0.4983 && 500 & 500 & (0, 9, 0, 0, 1) & 0.9997\\	             
&	             &&                             &                 && 1000 & 518 & (0, 9, 0, 0, 1) &  0.9997\\
&	             &&                             &                 && 10000 & 8643 & (0, 9, 0, 0, 1) &  0.9997\\
&	              &&                             &                 && 50000 &  49507 & (0, 9, 0, 0, 1) &  0.9997\\
&	              &&                             &                 &&  100000 &   49685         &   (0, 14, 0, 0, 1) & 0.9998\\\bottomrule
Parameters &	  	\multirow{2}{*}{$(n, m)$} && \multicolumn{2}{c}{Exhaustive search} && \multicolumn{4}{c}{Multivariate hypergeometric}\\\cline{4-5}\cline{7-10}
$(\beta, k)$ &	  && $\mathcal{R}^{*}$ & $\psi(\mathcal{R}^{*})$ && $n_{\tiny{\mbox{it}}}$  & $n_{\tiny{\mbox{ac}}}$ & $\mathcal{R}^{\bullet}$  & R.eff$_1$ \\\midrule	  
\multirow{5}{*}{(1, 1)} &	 (15, 5) && 	(0, 10, 0, 0, 0) & 0.4983 && 500 & 275 & (2, 6, 2, 0, 0) & 0.9996\\
&	              &&                                &                 && 1000 & 575 & (0, 5, 4, 0, 1) & 0.9996\\         
&  	              && 	                            &                 && 10000 & 5933 & (1, 1, 8, 0, 0) & 	0.9996\\              
&	              &&                                &                 && 50000 &  29809 & (1, 8, 0, 0, 1) & 0.9998\\
&	              &&                                &                 && 100000 &    26463      &       (0, 0, 6, 3, 6)                 & 0.9998\\\bottomrule
Parameters & \multirow{2}{*}{$(n, m)$} && \multicolumn{2}{c}{Exhaustive search} && \multicolumn{4}{c}{Multinomial}\\\cline{4-5}\cline{7-10}
$(\beta, k)$ & && $\mathcal{R}^{*}$ & $\psi(\mathcal{R}^{*})$ && $n_{\tiny{\mbox{it}}}$  & $n_{\tiny{\mbox{ac}}}$ & $\mathcal{R}^{\bullet}$ & R.eff$_1$ \\\midrule
\multirow{5}{*}{(2, 1)} & (20, 5) && 	(0, 15, 0, 0, 0) & 0.1113  &&  500 & 244 & (0, 0, 3, 10, 2) &   0.9998      \\        
&&&                             &                 && 1000 & 414 & (0, 0, 3, 10,2) & 0.9998\\
&&&                             &                 && 10000 & 4799 & (0, 9, 6, 0, 0) & 0.9998\\
&&&                             &                 && 50000 & 24105 & (0, 2, 13, 0, 0) & 0.9998\\
&&&                             &                 && 100000 & 45858 & (0, 12, 3, 0, 0) & 0.9998\\\bottomrule
Parameters & \multirow{2}{*}{$(n, m)$} && \multicolumn{2}{c}{Exhaustive search} && \multicolumn{4}{c}{Uniform}\\\cline{4-5}\cline{7-10}
$(\beta, k)$&&& $\mathcal{R}^{*}$ & $\psi(\mathcal{R}^{*})$ && $n_{\tiny{\mbox{it}}}$  & $n_{\tiny{\mbox{ac}}}$ & $\mathcal{R}^{\bullet}$  & R.eff$_1$ \\\midrule	             
\multirow{5}{*}{(2, 1)} &(20, 5) && 	(0, 15, 0, 0, 0) & 0.1113 && 500 & 440 & (0, 14, 0, 0, 1) & 0.9998\\	             
&&&                             &                 && 1000 & 987 & (0, 14, 0, 0, 1) &  0.9998\\
&&&                             &                 && 10000 & 8431 & (0, 14, 0, 0, 1) &  0.9998\\
&&&                             &                 && 50000 &  28759 & (0, 14, 0, 0, 1) &  0.9998\\
&&&                             &                 &&  100000 &   49685         &   (0, 14, 0, 0, 1) & 0.9998\\\bottomrule
Parameters & \multirow{2}{*}{$(n, m)$} && \multicolumn{2}{c}{Exhaustive search} && \multicolumn{4}{c}{Multivariate hypergeometric}\\\cline{4-5}\cline{7-10}
$(\beta, k)$ &&& $\mathcal{R}^{*}$ & $\psi(\mathcal{R}^{*})$ && $n_{\tiny{\mbox{it}}}$  & $n_{\tiny{\mbox{ac}}}$ & $\mathcal{R}^{\bullet}$  & R.eff$_1$ \\\midrule	  
\multirow{5}{*}{(2, 1)} &(20, 5) && 	(0, 15, 0, 0, 0) & 0.1113 && 500 & 246 & (4, 2, 0, 2, 7) & 0.9998\\
&&&                                &                 && 1000 & 581 & (0, 1, 6, 2, 6) & 0.9998\\         
&&& 	                            &                 && 10000 & 5387 & (0, 1, 4, 8, 2) & 	0.9998\\              
&&&                                &                 && 50000 &  26835 & (0, 3, 3, 6, 3) & 0.9998\\
&&&                                &                 && 100000 &    46463      &       (0, 0, 6, 3, 6)                 & 0.9998\\\bottomrule	              
	\end{tabular}
\end{table}

\begin{table}[h]
	\centering
	\scriptsize
	\caption{Comparison of near-optimal censoring schemes (for large $n$ and $m$) between the proposed approach and the VNS algorithm \citep{Ritwik_2016} under the criterion $\psi(\mathcal{R}) = \int_{0}^{1} \mbox{Var}[\ln \hat{X}_s]\mbox{d}s$ in the case of Weibull lifetime model.}
	\begin{tabular}{lllllllll}\toprule
		\multirow{2}{*}{$(n, m)$} && \multicolumn{2}{c}{VNS algorithm} && \multicolumn{2}{c}{Multinomial} &&  \multirow{2}{*}{R.eff$_2$}  \\ \cline{3-4}\cline{6-7}
		                                               &&  $\mathcal{R}^{\dagger}$ & $\psi(\mathcal{R}^{\dagger})$ & & $\mathcal{R}^{\bullet}$ & $\psi(\mathcal{R}^{\bullet})$ &&  \\\midrule
	(30, 10) && $(0^5, 20, 0^4)$ & 0.2826 && $(0^3, 1, 0, 2, 11, 3^2, 0)$ & 0.2826 && 1 \\  
	(30, 15) && $(0^7, 15, 0^7)$ & 0.2505 && $(0^5, 18, 2, 0^3)$ & 0.2826 &&  1\\  
	(35, 10) && $(0^6, 25, 0^3)$ & 0.2664 &&	$(0^4, 2, 5^2, 8, 4, 1)$																																																																																																																																																																																																																																																																																																																																																																																																																																																																																																																																																																			  & 0.2664 && 1 \\  
	(35, 15) && $(0^6, 20, 0^8)$ & 0.2342 && $(0^6, 2, 0, 1^2, 2, 0,5^2,4)$ & 0.2342 && 1 \\
	(35, 20) && $(0^6, 15, 0^{14})$ & 0.2189 && $(0^6, 4, 0, 2, 0^4, 2, 0, 2, 0, 2^2, 1 )$ & 0.2189 &&  1\\    
	(45, 10) && $(0^6, 35, 0^3)$ & 0.2442 && $(0^3, 4, 11, 4, 9, 2, 3, 2)$ & 0.2442 && 1 \\ 
	(45, 15) && $(0^7, 2, 0^6, 28)$ & 0.2118 && $(0^3, 1^2, 5^2, 0, 7, 0^2, 8, 2, 0, 1)$  & 0.2118 &&  1\\  
	(45, 20) && $(0^{19}, 25)$ & 0.1964 && $(0^5, 1, 0^2, 4, 1^2, 0, 1^2, 5, 1, 0, 2, 3, 5)$ & 0.1964 && 1 \\   	  \midrule
			\multirow{2}{*}{$(n, m)$} && \multicolumn{2}{c}{VNS algorithm} && \multicolumn{2}{c}{Uniform} &&  \multirow{2}{*}{R.eff$_2$}  \\ \cline{3-4}\cline{6-7}
&&  $\mathcal{R}^{\dagger}$ & $\psi(\mathcal{R}^{\dagger})$ & & $\mathcal{R}^{\bullet}$ & $\psi(\mathcal{R}^{\bullet})$ &&  \\\midrule	
	(30, 10) && $(0^5, 20, 0^4)$ & 0.2826 && $(0^4, 13, 6, 0^3,1)$ & 0.2826 && 1 \\
	(30, 15) &&  $(0^7, 15, 0^7)$ & 0.2505  && $(0^2, 5, 0, 4, 0, 1, 0^2, 3, 1, 0^3, 1)$ & 0.2505  && 1 \\ 
	(35, 10) &&  $(0^6, 25, 0^3)$ & 0.2664  && $(0, 1 7, 14, 0, 2, 0^2, 1)$	& 0.2664 && 1 \\   
(35, 15) && $(0^6, 20, 0^8)$ & 0.2342 && $(0^3, 15, 2, 1, 1, 0^7, 1)$ & 0.2342 && 1 \\ 
(35, 20) &&  $(0^6, 15, 0^{14})$ & 0.2189  && $(0^2, 1, 2, 7, 4, 0^{13},  1)$ & 0.2189 && 1 \\ 
(45, 10) &&  $(0^6, 35, 0^3)$ & 0.2442  && $(0, 1, 1,3 10, 3, 4, 3, 0^2, 1)$ & 0.2442 &&  1\\ 
(45, 15) &&  $(0^7, 2, 0^6, 28)$ & 0.2118  && $(0^2, 15, 5, 0, 2^2, 1, 4, 0^5, 1)$ & 0.2118 &&  1\\
(45, 20) &&  $(0^{19}, 25)$ & 0.1964  && $(0^2, 7, 14, 3, 0^{13}, 1 )$ & 0.1964 &&  1\\ 	  \midrule	    
 			\multirow{2}{*}{$(n, m)$} && \multicolumn{2}{c}{VNS algorithm} && \multicolumn{2}{c}{Multivariate hypergeometric} &&  \multirow{2}{*}{R.eff$_2$}  \\ \cline{3-4}\cline{6-7}
 &&  $\mathcal{R}^{\dagger}$ & $\psi(\mathcal{R}^{\dagger})$ & & $\mathcal{R}^{\bullet}$ & $\psi(\mathcal{R}^{\bullet})$ &&  \\\midrule	
 (30, 10) && $(0^5, 20, 0^4)$ & 0.2826 && $(0,2,0,1,3,1,1,6,1,5)$ & 0.2826 && 1 \\ 
 (30, 15) &&  $(0^7, 15, 0^7)$ & 0.2505  && $(0^3, 1, 0, 1, 0, 3, 0, 1, 0, 5, 2^2, 0)$ & 0.2505 && 1 \\  
 (35, 10) &&  $(0^6, 25, 0^3)$ & 0.2664  && $(1, 0, 5^2, 1, 2, 3^2, 5, 0)$ & 0.2664 &&  1\\   
 (35, 15) && $(0^6, 20, 0^8)$ & 0.2342 && $(0^4,1,0,3^2,2, 3,0,1,3,0,4)$ & 0.2342 && 1 \\  
 (35, 20) &&  $(0^6, 15, 0^{14})$ & 0.2189  && $(0^6, 1, 2, 5, 1^2, 0, 4, 1, 0^6)$ & 0.2189 && 1 \\ 
 (45, 10) &&  $(0^6, 35, 0^3)$ & 0.2442  && $(6, 2^2, 3, 4, 8, 4, 1, 4, 1)$ & 0.2442 && 1 \\ 
 (45, 15) &&  $(0^7, 2, 0^6, 28)$ & 0.2118  && $(2, 0^3, 1, 2, 1, 4, 1, 7, 0, 7, 1, 2^2)$ & 0.2118 && 1 \\
 (45, 20) &&  $(0^{19}, 25)$ & 0.1964  && $(2, 0^3, 1^2, 0^2, 1, 2, 0, 3, 0, 3, 2, 0, 3, 0, 2, 5)$ & 0.1964 && 1 \\\bottomrule                                       
	\end{tabular}

\end{table}

\section{Conclusions}
\paragraph{}
In this work, we have used a probability structure to develop a simple algorithm for the determination of optimal progressive censoring scheme. We have used three probability distributions for  $\pi_{\tiny{\mathcal{R}}}(\cdot)$ for illustrative purpose.   The comparisons made with a direct exhaustive search method and the VNS algorithm demonstrate the efficiency and good performance of the proposed method.

\section*{Appendix}
\subsection*{Variance and cost measures under Weibull model}
To derive explicit expressions for the two criteria used here, we first need the density of $X_{i:m:n}$, given by the Kamps-Cramer representation \citep[see][]{Bala_Pro_2007, Cramer_book}, as  
\begin{equation}\nonumber
	f_{X_{i:m:n}}(x; \theta) = \sigma_{i-1} \sum_{p=1}^{i} a_{p, i} \{1 - F(x; \theta)\}^{\gamma_{p}-1} f(x; \theta),~~\mbox{for}~~ i=1,\cdots, m,
\end{equation}where $\gamma_{r} = m - r + 1 + \sum_{i=r}^{m} R_i,$ for $r=1,\cdots, m$, $\sigma_{r-1}=\prod_{i=1}^{r} \gamma_{i}$, $a_{i,r} = \prod_{\substack{j=1\\ j\neq i}}^r \frac{1}{\gamma_j-\gamma_i}$ for $1\leq i \leq r \leq m$ with $a_{1, 1} = 1$, and $f(x; \theta)$ is the probability density function of $X$ with $\theta=(\beta, k)$. From the above density function of $X_{i:m:n}$, $E[X_{m:m:n}]$ can be found to be  
\begin{equation}\nonumber
	E[X_{m:m:n}] = \frac{1}{k} \Gamma\left(1+\frac{1}{\beta}\right) \sigma_{m-1}\sum_{p=1}^m \frac{a_{p,m}}{\gamma^{1+\frac{1}{\beta}}_p}.
\end{equation}

Finally, to derive an expression of the variance measure, we need the expression of Fisher information matrix about $\theta$ given by \citep[see][]{Ritwik_2016}

$$\mathcal{I}(\theta)=\begin{bmatrix}\mathcal{I}_{11}(\theta) && \mathcal{I}_{12}(\theta) \\ \mathcal{I}_{21}(\theta) && \mathcal{I}_{22}(\theta) \end{bmatrix},$$ where
\begin{eqnarray}\nonumber
	\mathcal{I}_{11}(\theta) &=& \frac{1}{\beta^2} \sum_{i=1}^m  \sum_{j=1}^i \sigma_{i-1} \frac{a_{j, i}}{\gamma_j} \int_0^{\infty} \left(  1+ \ln\left(\frac{z}{\gamma_j}\right)  \right)^2 e^{-z}\mbox{d}z, \\\nonumber
	\mathcal{I}_{22}(\theta) &=& \left( \frac{\beta}{k}  \right)^2 \sum_{i=1}^m  \sum_{j=1}^i \sigma_{i-1} \frac{a_{j, i}}{\gamma_j}, \\\nonumber
	\mathcal{I}_{12}(\theta)~ = ~\mathcal{I}_{21}(\theta) &=& \frac{1}{k} \sum_{i=1}^m  \sum_{j=1}^i \sigma_{i-1} \frac{a_{j, i}}{\gamma_j} \int_0^{\infty} \left(  1+ \ln\left(\frac{z}{\gamma_j}\right)  \right)e^{-z} \mbox{d}z.
\end{eqnarray}It is difficult to find the exact expression of $\int_0^1 \mbox{Var}[\ln \hat{X}_s]\mbox{d}s$. However, by using delta method, the asymptotic expression of $\mbox{Var}[\ln \hat{X}_s]$ can be found to be $\mbox{Var}[\ln \hat{X}_s]=(\mathcal{I}^{11}(\theta)/\beta^4)$ $(g(s))^2+(2\mathcal{I}^{12}(\theta)/\beta^2 k )g(s)+(\mathcal{I}^{22}(\theta)/k^2)$, where $g(s)=\ln\{ -\ln(1-s) \}$, and $\mathcal{I}^{11}(\theta)$, $\mathcal{I}^{12}(\theta)$ and $\mathcal{I}^{22}(\theta)$ are the elements of the inverse of the Fisher information matrix given by $$\mathcal{I}^{-1}(\theta)=\begin{bmatrix}\mathcal{I}^{11}(\theta) && \mathcal{I}^{12}(\theta) \\ \mathcal{I}^{21}(\theta) && \mathcal{I}^{22}(\theta) \end{bmatrix}.$$Therefore, an asymptotic expression of $\int_0^1 \mbox{Var}[\ln \hat{X}_s]\mbox{d}s$ can be obtained readily as
\begin{equation}\nonumber
	\frac{\mathcal{I}^{11}(\theta)}{\beta^4}\int_0^1 (g(s))^2 \mbox{d}s+\frac{2\mathcal{I}^{12}(\theta)}{\beta^2k}\int_0^1 g(s)\mbox{d}s+\frac{\mathcal{I}^{22}(\theta)}{k^2}.
\end{equation}

\bibliographystyle{apalike}
\bibliography{ritwik_ref}

\end{document}